# Innovation beyond intention: harnessing exaptation for technological breakthroughs

Youwei He[a, b, *], Jeong-Dong Lee[a, b], Seungmin Lee[a]

[a]Technology Management, Economics and Policy Program, Seoul National University, 1 Gwanak-ro, Gwanak-gu, 08826, Seoul, South Korea; [b]Integrated Major in Smart City Global Convergence, Seoul National University, 1 Gwanak-ro, Gwanak-gu, Seoul, 08826, South Korea

*Corresponding author

**Abstract**

The frameworks that explore scientific and technological evolution suggest that discoveries and inventions are intrinsic processes, while the wealth of knowledge accumulated over time enables researchers to make further advancements, echoing Newton's sentiment of "standing on the shoulders of giants." Despite the exponential growth in new scientific and technical knowledge, the consolidation–disruption (CD) index suggests a concerning decline in the disruptiveness of papers and patents. "Exaptation," a concept borrowed from biological evolution, is now recognized as a pivotal yet often neglected mechanism in technological evolution. Significant technologies often do not emerge out of thin air but rather result from the application of existing technologies in other domains. For instance, bird feathers initially served as waterproofing and insulation before enabling flight, and microwave ovens originated from radar magnetrons. Exaptation, acknowledged as the catalyst for "innovation beyond intention," signifies a cross-field evolutionary process that is driven by functional shifts in pre-existing knowledge, technology, or artifacts. In this study, we introduce the concept of exaptation value, deliberately excluding serendipity. Our analysis reveals that, despite a declining trend in the disruptiveness of innovation, there is an increasing trend in the application of cross-domain knowledge within the innovation process over time. We also explore the impact of technology exaptation on innovation disruptiveness and discuss how leveraging technology adaptability enhances innovation's disruptive potential.

# 1. Introduction

The exponential growth of innovation in the technology sector has been unmistakably evident in recent decades; however, recent studies suggest that new innovations are increasingly less likely to deviate from past trends in ways that steer technology in novel directions (Park et al., 2023). This rapid innovation with minimal technological change is attributable to shifts within technological domains, while technological evolution itself is discontinuous (Levinthal, 1998). Interdisciplinary research is championed as a catalyst for disruptive innovation, as it brings together diverse knowledge that can inject fresh perspectives into traditional research and become a pivotal source of creativity and innovation (Andriani & Cattani, 2016; Ganzaroli & Pilotti, 2010). "Exaptation" is a term from evolutionary biology that denotes a functional shift in a trait during evolution (Gould & Vrba, 1982). There are numerous examples of exaptation. The feathers of birds, initially evolved for thermal regulation and waterproofing, were later repurposed for flight. The concept of exaptation has extended into various other domains such as technology, economics, management, and pharmacy. Within the realm of technology, exaptation-driven innovation harnesses the latent functionalities of existing artifacts in new contexts (Beltagui et al., 2020). The growing disparity between the rise of exaptation and the declining rate of disruptive innovation (Collison & Nielsen, 2018; Horgan, 2015) necessitates an exploration of the interplay between exaptation-driven innovation and disruptive innovation.

Exaptation refers to an evolutionary discontinuity that arises from a functional shift in an existing trait or artifact (Andriani & Cattani, 2016). This process accelerates technological evolution, making it both rapid and discontinuous (Levinthal, 1998). Innovation driven by exaptation offers potential solutions to enduring questions about the genesis of novelty, especially in the context of radical innovation (Andriani & Cattani, 2016); however, the individual-level disruptive potential of exaptation remains underexplored despite a clearer understanding of how an ecosystem can foster disruption through an "exaptation seed" (Beltagui et al., 2020). The concept of creative destruction was introduced by Schumpeter (1942) and describes the disruption of existing technologies and the subsequent alteration of technological trajectories. The metaphor that disruptive works "stand on the shoulders of giants"—a sentiment echoed by Newton—can be traced back to Koyré (1952), Fleck (1981), and Popper (2014). Yet, through the lens of exaptation, these giants may hail from disparate fields. Disruptive innovations can emerge as a result of exaptation, especially since exaptation typically precedes disruption. Furthermore, the recognition of disruption is contingent upon its progeny, which remains uncertain before exaptation.

Inventions often arise from chance rather than intent. Serendipity has frequently played a role throughout the history of technological and scientific advancements (Vuong, 2022). Inherent uncertainty and the serendipitous nature of technological evolution make it impossible to predict all potential uses and applications of existing technologies in advance (Cattani & Mastrogiorgio, 2021). For example, the microwave oven originated from radar technology called magnetron (U.S. Patent 2,123,728). Raytheon scientist Percy Spencer obtained a patent for the microwave after noticing that radar emissions melted the candy in his pocket (U.S. Patent 2,495,429). In 1950, he secured the patent and commercialized the technology, marking a significant breakthrough (Mastrogiorgio & Gilsing, 2016). Serendipity encompasses not only scientific discoveries but also the identification of new applications for those discoveries, leading to exaptation (Garud et al., 2018). While exaptation is often associated with serendipity, distinguishing between the two can be challenging, as many instances involve a blend of both (Leporini et al., 2020). Technological exaptation is considered a key driver of "innovation beyond intention." The intersection of serendipity and exaptation is commonplace in the genesis of inventions, products, and other creative endeavors. In the pharmaceutical sector, medications initially approved for one disease frequently find off-label use for entirely different conditions (Andriani et al., 2017). Uncertainty in the exaptation process pertains to the successor, not to the innovation driven by exaptation, since such innovation has already undergone a functional shift from its predecessor.

In our research, we assess the degree of exaptation from the predecessors of the focal innovation, which differs from measuring serendipity in the successor as done in prior studies (Mastrogiorgio & Gilsing, 2016; Ferreira et al., 2020); furthermore, we characterize exaptation-driven innovation as innovation primarily sourced from exaptive knowledge. This extends beyond the mere assimilation of exaptive knowledge; the knowledge base must be substantially composed of exaptive knowledge. This distinction is crucial, because non-exaptation-driven innovation may also incorporate exaptive knowledge, but it does not constitute the source of the primary knowledge. In other words, interdisciplinary research qualifies as exaptation-driven innovation when innovations that create niches within their ecosystem incorporate knowledge from other fields.

We analyzed 3.4 million patents from 1980 to 2010, all of which contain citation information from the United States Patent and Trademark Office's (USPTO) Patents View database. This database encompasses over 35 million citations and 3.4 million abstracts. Leveraging this extensive dataset, we combined the consolidation–disruption (CD) index—a citation-based measure (Funk & Owen-Smith, 2017; Wu et al., 2019; Chen et al., 2021)—with a natural language processing (NLP) analysis (Reimers & Gurevych, 2019) of the abstracts. Our goal was to discern whether exaptation fosters disruptive innovation over time and across various fields. In the "Measurement of technology" section, we explore the use of the CD index method to evaluate the disruptiveness of patents. We also introduce a novel approach for quantifying exaptation by examining content similarity and field distance and formulate three research questions to probe the link between exaptation-driven innovation and disruptiveness. To illustrate the enabling influence of exaptation on disruptiveness, we conducted three case studies, similar to those presented by Funk and Owen-Smith (2017). Our findings were further substantiated by validating the association between exaptation and disruptiveness through a regression model, conducting robustness tests, and discussing the implications of our results.

## 2. Measurement of technology

### 2.1 Measurement of disruptiveness

The CD index measurement (Funk & Owen-Smith, 2017) was introduced to gauge the disruptive impact of science and technology through citation networks. The core principle of the CD index hinges on whether a focal work's descendants also cite its predecessor. Disruptive works' progeny may not consistently acknowledge their antecedents, while the successors of consolidating works are more inclined to reference the focal work's ancestors. The CD index scale spans from −1 (consolidating) to 1 (destabilizing). This measure has been refined and expanded in subsequent research (Balachandran & Hernandez, 2018; Fortunato et al., 2018; Jia et al., 2019; Li & Chen, 2022; Ruan et al., 2021; Wei et al., 2023; Wu et al., 2019). We evaluate disruption within patent data using the disruption (*D*) value as defined by Wu et al. (2019). Employing Equation (1), which delineates the three types of descendants as illustrated in Fig. 1, we find that the *D* value aligns precisely with the CD index.

$$CD = D = \frac{n_i - n_j}{n_i + n_j + n_k} \qquad (1)$$

Where $n_i$ refers to the count of descendants that cite the focal work; $n_j$ refers to the count of descendants that cite the focal work and the focal work's ancestor; and $n_k$ refers to the count of descendants that cite the focal work's ancestor but not the focal work. To quantify the long-term variations in the CD value, we calculate the difference between the 10-year CD index and the 5-year CD index. This difference is used to indicate growth in disruptiveness, as expressed in Equation (2):

$$\Delta CD = CD_{10} - CD_5, \tag{2}$$

Where $\Delta CD$ refers to the long-term change in CD value, $CD_{10}$ refers to the 10-year CD index, and $CD_5$ refers to the 5-year CD index.

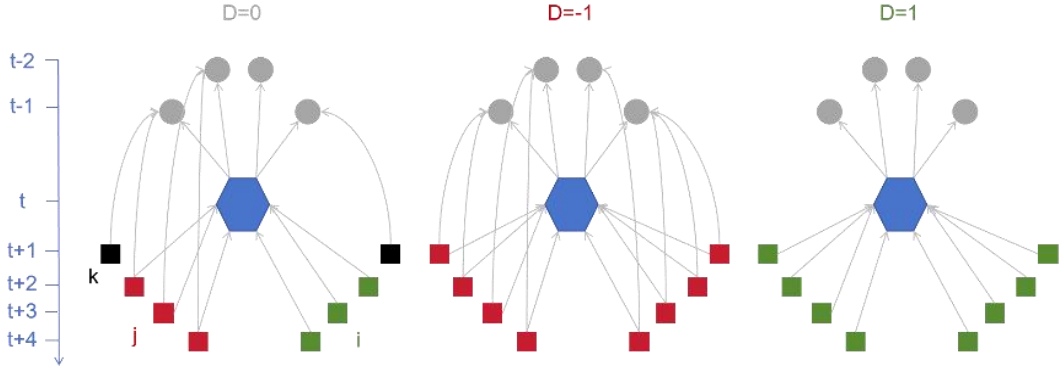

**Fig. 1** The approach of quantifying disruption

*Note*: The figure provides a simplified depiction of disruption within citation networks. It consists of three distinct networks, each featuring focal papers (blue hexagons), their references (gray circles), and subsequent works (squares). Subsequent works may reference the focal paper alone ("i" in green), both the focal paper and its references ("j" in red), or solely its references ("k" in black). A patent may be disrupting (D = 1), neutral (D = 0) or developing (D = −1).

2.2 Measurement of exaptation

Technology exaptation can catalyze disruptive innovation by cultivating ecosystems and niches from existing technologies in other domains (Andriani & Cohen, 2013; Beltagui et al., 2020). Exaptation often embodies uncertainty, because original inventions are not anticipated to be repurposed by future inventions (Bonifati, 2010). Thus, prior studies assessed technological exaptation based on uncertain progeny at the time of creation (Ferreira et al., 2020; Mastrogiorgio & Gilsing, 2016). However, from the perspective of exaptation-driven innovation, exaptation appears definitive when analyzing the impact of ancestral technologies on a focal patent rather than on its descendants. Invention has been described as a process of recombinant search across a technological terrain (Fleming & Sorenson, 2001), suggesting that invention is an amalgamation of specific technologies that can be adapted for alternative uses (Arthur, 2010). Innovation emerges from its predecessors through two types of changes: the alteration of the technology's content and the shift in the domain where the technology is applied (Norman & Verganti, 2014).

Our research quantifies exaptation by the changes in the content and application field within the innovation. The more similar the content of the focal patent is to its ancestor, the higher the exaptation value; conversely, the more distinct the application field of the focal patent is from its ancestor, the higher the exaptation value. We embed all the patent abstracts into 384-dimensional vectors using a sentence transformer (Reimers & Gurevych, 2019), a pre-trained NLP model, and calculate the cosine similarity between each patent and its ancestors based on their vectors to determine content similarity.

The Cooperative Patent Classification (CPC) code signifies the technical field of a patent, and a patent may have multiple CPC codes. We compute the average Jaccard similarity of patents and their ancestors using their four-level CPC codes (e.g., CPC section: "A"; CPC class: "A63"; CPC subclass: "A63B"; and CPC group: "A63B71/146") to assess field similarity. Field distance is defined as one minus the field similarity. This field distance metric aligns with the functional distance measure and quantifies the divergence between the initial application and subsequent emergent uses within the pharmaceutical sector (Andriani et al., 2017; Park, 2021). We utilize the product of content similarity and field distance to express the exaptation value in Equation (3). By evaluating the exaptation value, we can determine the degree of exaptive knowledge that the focal patent has inherited from its ancestor.

$$E_{i,k} = CS_{i,k} * FD_{i,k} \tag{3}$$

Where $i$ is the focal patent, and $k$ is an ancestor of the focal patent. $E_{i,k}$ refers to the exaptation value between focal patent $i$ and its ancestor $k$. $CS_{i,k}$ refers to the content similarity between focal patent $i$ and its ancestor $k$. $FD_{i,k}$ refers to the field distance between focal patent $i$ and its ancestor $k$. Ancestor refers to the patent cited by the focal patent.

Numerous distinct exaptation values can be associated with one patent, because a single patent can have multiple ancestors. We define the maximum exaptation value of patent $i$ as $E_{max}(i)$ in Equation (4), the minimum exaptation value of patent $i$ as $E_{min}(i)$ in Equation (5), and the average exaptation value of patent $i$ as $E_{avg}(i)$ in Equation (6). The maximum exaptation value signifies the utmost degree to which the focal patent strives to assimilate insights from its ancestors via exaptive learning (Pauleen, 2017); conversely, the minimum exaptation value delineates the baseline of such exaptive learning processes. By establishing the baseline exaptation value, we can define a threshold to determine whether an innovation is driven by exaptation. Should the minimum exaptation value exceed this threshold, it indicates that all the measured exaptation values surpass it, confirming the innovation as an exaptation-driven innovation. The average exaptation value conveys the mean level of exaptive knowledge acquired from its ancestral patents. An overview of the measurement approach for CD index and the maximum exaptation value, along with the distribution of $CD_5$ and $E_{max}$, based on USPTO data, is shown in Fig. 2.

$$E_{max}(i) = CS_{i,k^+} * FD_{i,k^+}, \tag{4}$$
$$k^+ = argmax_{k \in X}(CS_{i,k} * FD_{i,k})$$

Where $k^+$ refers to the ancestor which induce the maximum exaptation value to the patent $i$.

$$E_{min}(i) = CS_{i,k^-} * FD_{i,k^-}, \tag{5}$$
$$k^- = argmin_{k \in X}(CS_{i,k} * FD_{i,k})$$

Where $k^-$ refers to the ancestor which induce the minimum exaptation value to the patent $i$.

$$E_{avg}(i) = \frac{1}{m}\sum_{k=1}^{m} CS_{i,k} * FD_{i,k}, \tag{6}$$

Where $m$ refers to the ancestor count, patent $i$ has.

Field distance dispersion, as defined in Equation (7), represents the variance between the maximum and minimum field distances associated with the focal patent. This metric quantifies the extent of field variation inherited from its predecessors. A larger field distance dispersion suggests that the focal patent has a wider absorptive capacity across various fields. Content similarity dispersion, outlined in Equation (8), measures the range between the highest and lowest content similarities linked to the focal patent. This metric assesses the breadth of technological knowledge diversity derived from its antecedents. Greater content similarity dispersion implies that the focal patent is versatile and functions across a wider array of technologies.

$$FD_{disp}(i) = FD_{max}(i) - FD_{min}(i), \qquad (7)$$

Where $i$ refers to the focal node; $FD_{max}(i)$ refers to the maximum field distance from focal node's ancestor to the focal node; $FD_{min}(i)$ refers to the minimum field distance from focal node's ancestor to the focal node; and $FD_{disp}(i)$ refers to the field distance dispersion.

$$CS_{disp}(i) = CS_{max}(i) - CS_{min}(i), \qquad (8)$$

Where $i$ refers to the focal node; $CS_{max}(i)$ refers to the maximum content similarity from focal node's ancestor to the focal node; $CS_{min}(i)$ refers to the minimum content similarity from focal node's ancestor to the focal node; and $CS_{disp}(i)$ refers to the content similarity dispersion.

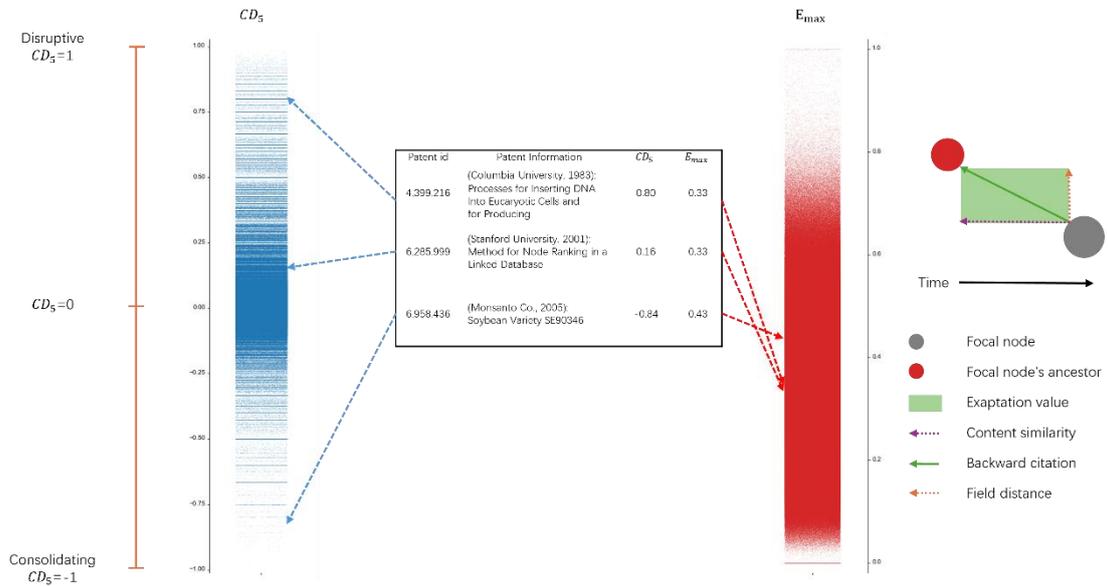

**Fig. 2** Overview of the measurement approach

*Note*: This figure presents a schematic representation of the exaptation value. At the center, a table showcases the three patent examples provided by Funk and Owen-Smith (2017), each with their respective 5-year CD index ($CD_5$) and maximum exaptation values ($E_{max}$). To the left and right of the table are blue and red bars, symbolizing the distribution of $CD_5$ and $E_{max}$ values, respectively, for a dataset of patents (n=3,420,673) issued by the USPTO from 1980 to 2010. The table highlights sample patents alongside pertinent data. The vertical axis of each bar correlates to the $CD_5$ and $E_{max}$ values, while the horizontal axis is designed to reduce point overlap. Darker shades within each vertical strip denote areas of higher density in the distribution, indicating more frequently observed $CD_5$ and $E_{max}$ values. For an in-depth look at the distribution of $CD_5$ and $E_{max}$, refer to Extended Data Fig. 1 in the supplementary material. An illustrative computation of the exaptation value derived from the focal node's ancestor is depicted on the far right.

## 3. Research questions

### 3.1 Disruption decreases but exaptation increases over time

The average CD index displays a downward trend over time, as shown in Fig. 3a. This trend suggests that patents are becoming increasingly conservative, adhering more closely to the established trajectory set by their predecessors and consequently less likely to propel science and technology into uncharted territories. Efforts to implement exaptive knowledge extracted from ancestral patents reflect the widespread dissemination of expertise across a multitude of fields. Inventors are not merely concentrating on leveraging the foundations established by their predecessors; they are also drawing on knowledge from various domains to fuel exploration, which could result in significant technological innovations. Over time, this exchange of knowledge becomes increasingly ubiquitous and profound. In contrast to the CD index, the average maximum exaptation value has an ascending trend, as illustrated in Fig. 3b. This trend is encouraging, as it signifies that innovations are more frequently integrating knowledge from diverse fields. The integration of a vast amount of knowledge from diverse fields has contributed to the growing "burden of knowledge" as science advances (Jones, 2009). The decline in disruptive innovations and increase in knowledge absorption from other fields are not contradictory. A study by Linzhuo et al. (2020) found that 21st-century physics empirically demonstrates that dense, centralized collaboration is associated with a reduction in the space of ideas.

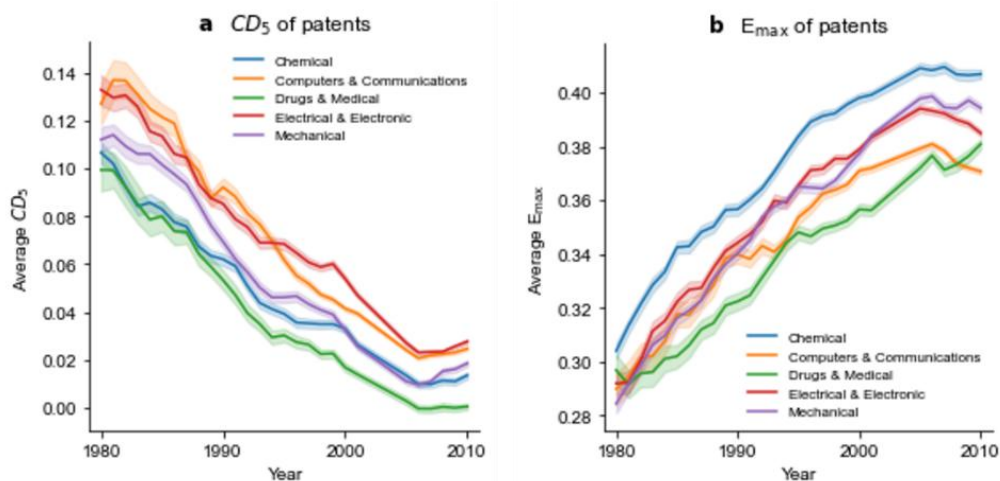

**Fig. 3** Decline of disruptive technology and incline of maximum exaptation value

*Note:* **a**, Decline in $CD_5$ over time for patents (n=3,420,673). For patents, lines correspond to the National Bureau of Economic Research (NBER) technology categories; from 1980 to 2010, the magnitude of the decline in $CD_5$ ranges from 79.2% (Electrical and electronic) to 99.6% (Drugs and medical). **b**, Incline in $E_{max}$ over time for patents (n=3,420,673). The magnitude of the incline in $E_{max}$ ranges from 27.9% (Computers and communications) to 38.6% (Mechanical). Shaded bands correspond to 95% confidence intervals.

## 3..2 Highly exaptive patents may not be considered disruptive at first

The maximum exaptation value denotes the transference of knowledge from diverse fields with minimal modifications. It essentially signifies the repurposing of similar technologies in different sectors. Although applications across fields have the potential to be groundbreaking, simple replication with minor adjustments may not lead to disruptive innovation. For instance, in 1958, Ford's Nucleon emerged as the pioneering nuclear-powered concept car; its propulsion technology was inspired by fission-driven steam engines used in nuclear submarines. Yet, the car did not achieve commercial success. Its downfall was not due to integrating the radioactive core; rather, it was the challenge of managing the immense energy released by such a reactor on an automotive scale (Gilboy, 2021). In other words, a patent with high exaptation but not a good integration may not necessarily be disruptive. Innovations that effectively apply and synthesize knowledge from other fields are more likely to be disruptive. The metrics of field distance and content similarity dispersion act as barometers for the synergistic fusion of exaptive knowledge in the innovation process. In their study, Rothaermel et al. (2010) warned of the negative consequences arising from the merger of incompatible innovations. Thus, increased levels of field distance or content similarity dispersion may lead to less disruptive innovations.

    The scatter plots in Fig. 4b reveal no clear linear relationship between the maximum exaptation values and the $CD_5$ within the CPC class level on average, suggesting that patents with high exaptation do not consistently equate to disruptiveness. In contrast, a clear linear correlation exists between the minimum exaptation values and the $CD_5$ (Fig. 4a), indicating that a higher baseline of exaptation values might enhance disruptiveness. There are also significant negative correlations between the dispersion of field distance and the $CD_5$ (Fig. 4c), as well as between the dispersion of content similarity and $CD_5$ (Fig. 4d). These findings imply that integrating knowledge from sources with vastly different technological content or fields can lead to reduced disruptiveness.

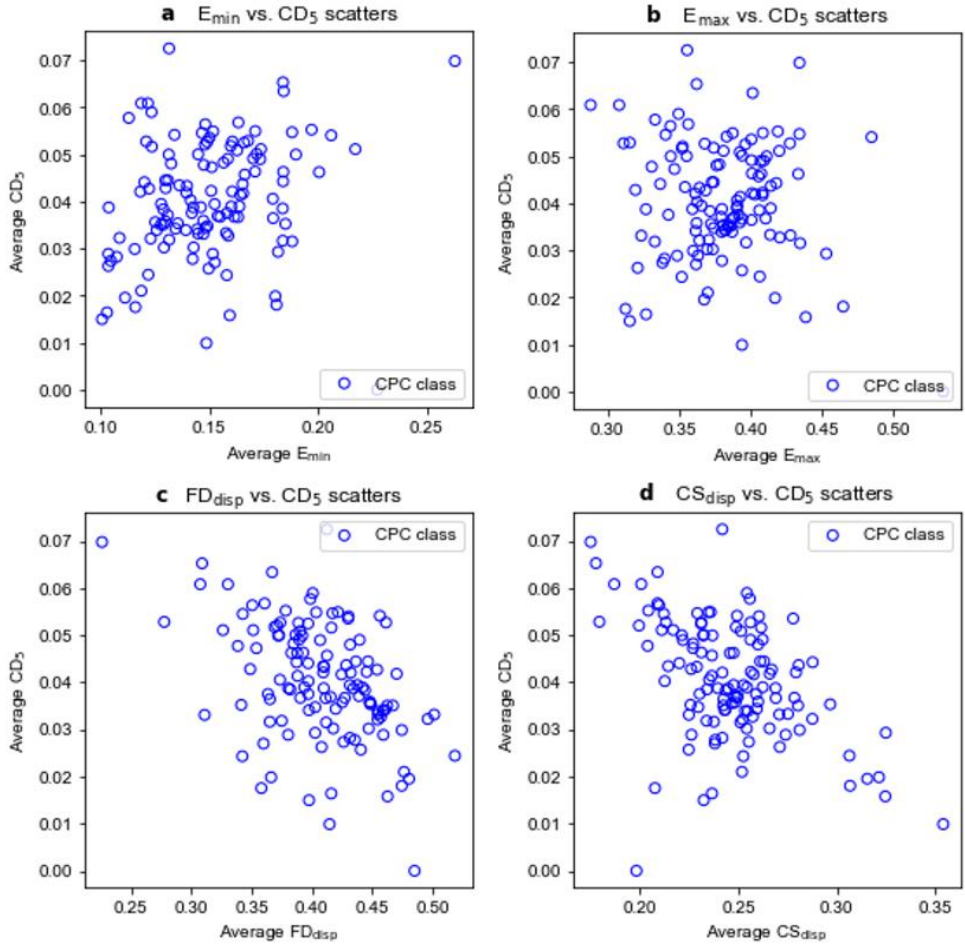

**Fig. 4** The scatters of exaptation values and CD index

*Note:* **a**, for USPTO patents from 1980 to 2010 (n=3,420,673); the scatter plot shows the scatters of average $CD_5$, and average $E_{min}$ values classified in 124 CPC classes. **b**, the scatter plot shows the scatters of average $CD_5$ and average $E_{max}$ values classified in 124 CPC classes. **c**, the scatter plot shows the scatters of average $CD_5$ and average $FD_{disp}$ values classified in 124 CPC classes. **d**, the scatter plot shows the scatters of average $CD_5$ and average $CS_{disp}$ values classified in 124 CPC classes.

### 3.3 Highly exaptive patents have the potential for continued disruptive growth

The $CD_5$ value is derived from the count of its various types of descendants and is subject to change over time. Some patents exhibit an increasing CD index trend, while others show a decreasing pattern. Certain patents, akin to the "sleeping beauties" described by Anthony (2015), maintain a steady $CD_5$ value for a period before experiencing a sudden surge. By assessing the change in CD values (ΔCD) using Equation (2), we can investigate the interplay between the evolution of disruptiveness and exaptation, including the "sleeping beauties," which are very easy to neglect. The scatter plots do not reveal a definitive linear correlation between the maximum exaptation value and the average ΔCD at the CPC class level (Fig. 5b); however, there is a notable positive association between the average minimum exaptation and the average ΔCD (Fig. 5a), suggesting that patents with a higher baseline of exaptation values may foster increased disruptiveness over time. Additionally, significant negative correlations are observed both between the dispersion of field distance and the rise in ΔCD, and between the dispersion of content similarity and the ΔCD (Figs. 5c and 5d). These correlations imply that the extent of field distance and content similarity dispersion may influence the disruptive potential of patents in the long term.

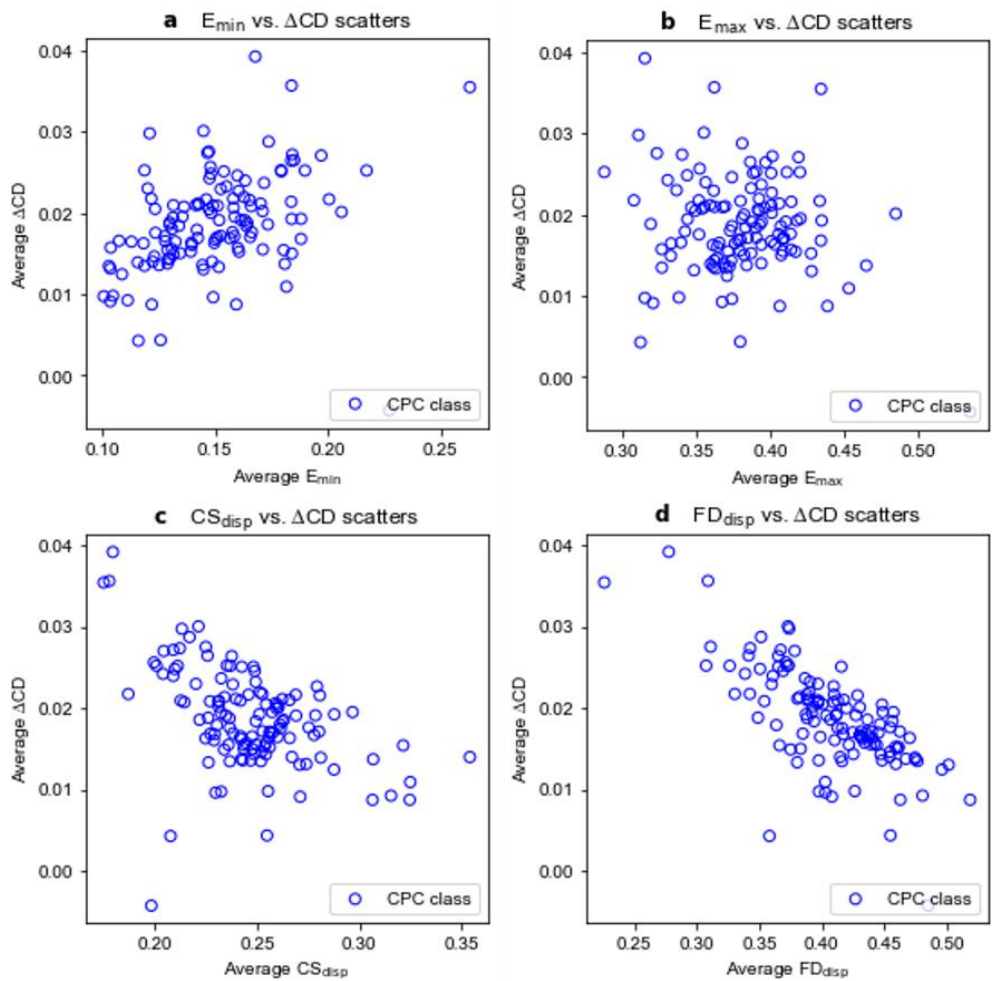

**Fig. 5** Scatter plots of exaptation values and changes of CD index

*Note*: **a**, for USPTO patents from 1980 to 2010 (n=3,420,673), the scatter plot shows the scatters of average ΔCD, and average $E_{min}$ values classified in 124 CPC classes. **b**, the scatter plot shows the scatters of average ΔCD and average $E_{max}$ values classified in 124 CPC classes. **c**, the scatter plot shows the scatters of average ΔCD and average $FD_{disp}$ values classified in 124 CPC classes. **d**, the scatter plot shows the scatters of average ΔCD and average $CS_{disp}$ values classified in 124 CPC classes.

## 4. Methods

### 4.1 Descriptive statistics and correlations

We examine the correlation between exaptation values and the CD index using data on the 3.4 million U.S. utility patents issued between 1980 and 2010, as established in prior research (Park et al., 2023). Our analysis omits patents without ancestral links. Due to differing preprocessing standards for patents and citations, the CD index values may vary from those reported in other studies. The fundamental covariates and the computation methodology for the CD index are derived from earlier studies on the patent network (Funk & Owen-Smith, 2017; Park et al., 2023; Wu et al., 2018).

Table 1 outlines the descriptive statistics and correlations. The values of the $CD_5$ index approximate a normal distribution following a logarithmic transformation, as illustrated in Extended Data Fig. 1a in the supplementary material, with a mean of 0.04, a standard deviation of 0.14, and a range from −1 to 1. The distribution is slightly skewed toward the right, which suggests that inventions with moderate destabilization are more prevalent than those with minor consolidation, echoing the results of previous studies (Funk & Owen-Smith, 2017). The maximum exaptation ($E_{max}$) value also conforms to a normal distribution, as shown in Extended Data Fig. 1b, with a mean of 0.37, a standard deviation of 0.14, and a range from 0 to 1. Despite a negative correlation between $E_{max}$ and both $CD_5$ and $\Delta CD$, a positive relationship is maintained when controlling for field distance dispersion and content similarity dispersion in the regression analysis. The minimum exaptation ($E_{min}$) and average exaptation ($E_{avg}$) values are positively correlated with both the $CD_5$ and the $\Delta CD$. $E_{max}$, $E_{min}$, and $E_{avg}$ exhibit positive correlations with one another. $E_{max}$ represents the maximal effort in assimilating exaptive knowledge, whereas $E_{min}$ denotes the foundational level of exaptation-driven innovation. This suggests that intensified efforts to integrate exaptive knowledge may enhance the probability of achieving exaptation-driven innovation; however, there is a negative correlation between $CD_5$ and $\Delta CD$, suggesting that disruptiveness may decline over time.

Both $CS_{disp}$ and $FD_{disp}$ exhibit a negative correlation with the $CD_5$ and the $\Delta CD$, indicating that disruptive innovation would concentrate on either field exploration or content exploitation to a certain extent, to avoid becoming a chaotic hodgepodge. In contrast, both the $CS_{disp}$ and $FD_{disp}$ show a positive correlation with the $Impact_5$, which shows the forward citation count in 5 years of the focal patent. This suggests that innovations that draw on a diverse range of knowledge and technologies from their predecessors are likely to yield a greater number of subsequent successors. The positive correlation between ancestor count (backward citation count of the focal patent) and $Impact_5$ substantiates this finding. In contrast, the negative correlation between ancestor count and $CD_5$ indicates disruptiveness related to less ancestor count (Funk & Owen-Smith, 2017). We use five control variables in the regression models: "Mean age of team member," "Mean age of work cited," "Ratio of self-citations to total work cited," "Dispersion in age of work cited," and "Mean number of prior works produced by team members." The variables are all negatively related to the $CD_5$ and $\Delta CD$.

**Table 1** Descriptive statistics and correlations

| Variables | Mean | SD | 1 | 2 | 3 | 4 | 5 | 6 | 7 | 8 | 9 |
|---|---|---|---|---|---|---|---|---|---|---|---|
| 1. $CD_5$ | 0.042 | 0.14 | 1 | | | | | | | | |
| 2. $\Delta CD$ | 0.017 | 0.098 | -0.164 | 1 | | | | | | | |
| 3. $E_{max}$ | 0.369 | 0.143 | -0.148 | -0.084 | 1 | | | | | | |
| 4. $E_{min}$ | 0.133 | 0.104 | 0.192 | 0.106 | 0.219 | 1 | | | | | |
| 5. $E_{average}$ | 0.241 | 0.1 | 0.021 | 0.01 | 0.755 | 0.72 | 1 | | | | |
| 6. $CS_{disp}$ | 0.259 | 0.182 | -0.268 | -0.147 | 0.404 | -0.42 | 0.016 | 1 | | | |
| 7. $FD_{disp}$ | 0.426 | 0.294 | -0.254 | -0.142 | 0.533 | -0.537 | 0.025 | 0.624 | 1 | | |
| 8. $Impact_5$ | 3.352 | 5.947 | 0.05 | -0.05 | 0.076 | -0.074 | -0.005 | 0.135 | 0.108 | 1 | |
| 9. Ancestor count | 10.376 | 22.416 | -0.117 | -0.065 | 0.244 | -0.2 | 0.014 | 0.428 | 0.324 | 0.214 | 1 |
| 10. Mean age of team members | 5.792 | 5.938 | -0.079 | -0.044 | 0.094 | -0.078 | 0.012 | 0.177 | 0.124 | 0.01 | 0.119 |
| 11. Mean age of work cited | 7.313 | 4.1 | -0.082 | -0.034 | 0.163 | -0.059 | 0.069 | 0.194 | 0.183 | -0.09 | 0.165 |
| 12. Ratio of self-citations to total work cited | 0.071 | 0.18 | -0.053 | -0.008 | -0.047 | -0.023 | -0.036 | 0.04 | -0.041 | -0.019 | -0.031 |
| 13. Dispersion in age of work cited | 9.84 | 7.883 | -0.256 | -0.138 | 0.379 | -0.291 | 0.062 | 0.558 | 0.521 | 0.06 | 0.406 |
| 14. Mean number of prior works produced by team members | 11.062 | 54.583 | -0.025 | -0.015 | 0.04 | -0.04 | 0.006 | 0.086 | 0.054 | 0.024 | 0.064 |

| Variables | 10 | 11 | 12 | 13 | 14 |
|---|---|---|---|---|---|
| 10. Mean age of team members | 1 | | | | |
| 11. Mean age of work cited | 0.199 | 1 | | | |
| 12. Ratio of self-citations to total work cited | 0.142 | -0.134 | 1 | | |
| 13. Dispersion in age of work cited | 0.219 | 0.577 | -0.086 | 1 | |
| 14. Mean number of prior works produced by team members | 0.225 | 0.028 | 0.078 | 0.059 | 1 |

4.2 Case studies

The examples featured in Extended Data Table 1 in the supplementary material span a variety of influential innovations across diverse industries and technological domains, as chronicled in the seminal study by Funk and Owen-Smith (2017). These instances elucidate the relationship between the CD index and exaptation, as well as the degree of interconnected within each classification. Next, we examine the three distinct innovations listed in the table of Fig. 2—glyphosate-resistant soybeans, a method for ranking online search results, and a eukaryotic co-transformation technique—each positioned at unique junctures on the continuum of disruptiveness and exaptiveness.

Consider Monsanto's Patent 6,958,436, titled "Soybean variety SE90346." This patent is a prime example of a consolidated invention. It outlines a genetically modified soybean variety that is resistant to glyphosate, an herbicide patented by Monsanto in the 1970s. Glyphosate, the key ingredient in Monsanto's top-selling herbicide Roundup, is enhanced by the soybean's additional features: increased yield, disease resistance, and shatter resistance, all achieved through Monsanto's exclusive genetic modifications. The genetically engineered soybeans enhance the utility of previously patented chemical and biological technologies by broadening their applications and limiting competition (Graff et al., 2003; Pollack, 2009).

At the time of the patent application, Monsanto had acquired nearly all the companies that owned the technologies cited in the patent, with all the referenced technologies exhibiting over 0.92 content similarity with the focal patent. This reflects the strategic intent of the established firms to increase the value of their knowledge assets, as posited by innovation theory (Sosa, 2011). It also provides valuable insights into corporate strategies concerning litigation over competitors' intellectual property. The focal patent references five ancestral patents issued between 1992 and 1998. The ancestral Patent 5,084,082—the only one owned by DuPont and granted in 1992—shows a field distance of 0.25 and a content similarity of 0.4 with the focal patent, leading to an exaptation value of 0.1. The subsequent ancestral patents all have content similarities above 0.92. Patent 5,304,728—granted in 1994 with the smallest field distance of 0—has the lowest exaptation value of 0, which is below the average reported in Extended Data Table 1. The other three patents have a field distance of 0.46. The focal patent's content similarity and field distance dispersions are 0.53 and 0.46, respectively, both exceeding the averages in Table 1. The highest expansion value, 0.43, comes from Patent 5,767,350, issued in 1998. The significant content similarity dispersion, combined with the low expectation values, results in a $CD_5$ of -0.84, which is indicative of consolidating innovation.

In this section, we delve into Patent 6,285,999—issued in 2001 and entitled "Method for node ranking in a linked database." The patent, known as PageRank, is a fundamental algorithm Google employs to assess the relevance of web pages. Prior to the advent of PageRank, search engines relied predominantly on suboptimal methods to rank search outcomes (Brin & Page, 1998). Owned by Stanford University and exclusively licensed to Google, PageRank introduced an innovative approach based on social network theory principles, which evaluates web pages according to the volume of inbound links from other websites. The patent in focus references seven preceding patents issued from 1990 to 2000. In contrast to the Monsanto scenario, the content similarities range from 0.31 to 0.67, averaging 0.49, and the field distances span 0.21 to 0.75, with an average of 0.47. These figures suggest that the focal patent significantly diverges from its predecessors in terms of technical content and functional distance. Nonetheless, the exaptation values lie between 0.10 and 0.33, which are not comparatively high. The dispersions in content similarity and field distance are 0.36 and 0.54, respectively. The $CD_5$ value is 0.16, indicative of a disruptive innovation, albeit with a moderate $CD_5$ rating of disruptive innovation.

In our final analysis, we scrutinize the seminal Axel Patent 4,399,216, which was granted in 1983 and heralded a new era in eukaryotic co transformation. This patent is not merely an innovation but a keystone of modern biotechnology, pivotal in spawning a plethora of contemporary protein-based therapeutics expressed through eukaryotic vectors (Center, 2003). Its significance was also recognized by Columbia University, which licensed it to over thirty entities (Dudzinski, 2005). The patent outlines a transformative method for the insertion of exogenous genes into cellular frameworks, culminating in the synthesis of the associated proteins. Initially, this patent was connected to two antecedents. Nevertheless, the one from 1974 was disregarded for lacking an abstract. The surviving ancestral Patent 4,195,125, is characterized by a content similarity of 0.37 and a field distance of 0.88, culminating in an exaptation value of 0.33—remarkably high relative to its peers. With only a single antecedent retained, the dispersion metrics for both content similarity and field distance plummet to zero, indicating an absence of variance and an extraordinary $CD_5$ index of 0.8, underscoring its exceptional stature.

4.3 Regression analysis

Our study investigates the relationship between disruptiveness and three exaptation metrics—$E_{max}$, $E_{min}$, and $E_{avg}$—while controlling for prior knowledge using ordinary least squares (OLS) regression models. These models adhere to the methodology established by Park et al. (2023) and are employed to forecast the $CD_5$ index for individual patents. We consider five indicators of prior knowledge utilization: content similarity dispersion ($CS_{disp}$), field distance dispersion ($FD_{disp}$), the ratio of self-citations to total work cited, the mean age of cited works, and the dispersion in age of work cited. The latter metric reflects the span of grant years among the referenced patents. Each variable is defined at a patent level.

The $CS_{disp}$ and $FD_{disp}$ quantify the diversity of technology content and application fields among the cited works. The self-citation ratio is calculated as the fraction of cited works that include at least one inventor in common with the focal patent relative to the total number of citations. To counteract potential confounding factors, we integrate both year and field fixed effects into our models. Year fixed effects account for temporal influences that uniformly affect all patents, such as overarching technological trends. Field fixed effects control for sector-specific constants that are stable over time, such as the inherent appreciation of disruptive innovations within certain domains. Our regression models offer greater precision than descriptive visualizations and adjust for field variations by using 39 National Bureau of Economic Research (NBER) technology subcategories. These categories encompass fields such as "agriculture," "food," "textiles," "coatings," "gas," 'organics," and "resins," granting a detailed perspective within the overarching chemistry technology category. This also allows for a more nuanced understanding of the impact of exaptation on disruptiveness across technological domains.

We incorporate controls for the "mean age of team members"—also known as "career age," which is the difference between the grant year of the focal patent and the year in which each inventor was first granted a patent—and the "mean number of prior works produced by team members." While an uptick in self-citation rates might suggest inventors are increasingly focusing on their own previous work, this trend could also be influenced by the sheer volume of prior work available for citation. Similarly, an increase in the age of work cited could imply that inventors are not keeping pace with current developments; however, this may also be a consequence of the aging workforce in the technology sector (Blau et al., 2017; Cui et al., 2022). For instance, more seasoned inventors might have a greater affinity for or give more consideration to earlier works, or they might be more inclined to resist new changes (Azoulay, 2019). By including these control variables, we aim to account for these alternative factors.

Our analysis reveals that harnessing prior knowledge, characterized by more exaptation technology, reduced diversity in technology content and fields, lesser reliance on one's own past work, and older references correlates with the creation of more disruptive technologies. This association persists even when considering the average age of team members, the span of the grant years of cited patents, and the number of prior works produced by the team. These insights are derived from our regression analyses, as detailed in Table 2. Models 1, 3, and 5 present comprehensive regression models that include fixed effects for year and field, each utilizing different exaptation measures ($E_{max}$, $E_{min}$, and $E_{avg}$); Models 2, 3, and 6 are without absorbing both fixed effects.

Our findings reveal a consistent pattern in technology: the exaptation coefficients are uniformly positive and statistically significant for patents ($E_{min}$: 0.074, $P < 0.001$; $E_{max}$: 0.009, $P < 0.001$; $E_{avg}$: 0.040, $P < 0.001$). This suggests that patents with higher exaptation values tend to be more disruptive. Notably, $E_{min}$ exhibits the most substantial influence on disruption among the exaptation values, with its coefficient exceeding 0.07. When all the other variables are held constant at their means, the predicted $CD_5$ for patents increases by 18.3% with a one standard deviation increase in $E_{min}$. Moreover, both the content similarity dispersion (−0.073 to −0.079, $P < 0.001$) and field distance dispersion (−0.037 to −0.051, $P < 0.001$) show negative coefficients and are statistically significant, indicating that less diversity in the technological content and fields of prior art is associated with greater disruption.

With all other variables constant, the predicted CD5 for patents decreases by 31.5% to 34.3% with a one standard deviation increase in content similarity dispersion, and by 26% to 35.7% with a one standard deviation increase in field distance dispersion. The ratio of self-citations to total citations has a negative and significant coefficient for patents (−0.041 to −0.042, $P < 0.001$), implying that patents relying more on the inventor's previous work tend to be less disruptive. Keeping all other variables constant, the predicted $CD_5$ for patents decreases by 17.5% to 18.1% with a one standard deviation increase in this ratio. The interaction coefficient between the mean age of cited work and the dispersion in the age of cited work is positive and significant for patents (0.0001, $P < 0.01$), suggesting that when the dispersion in the age of cited work is held constant, engaging with older prior art correlates with higher disruption. The predicted $CD_5$ for patents increases by 18.5% to 19.5% with a one standard deviation increase in the mean age of cited work, again with all other variables held constant. In summary, our regression analysis suggests that exaptation fosters disruptiveness. Conversely, the employment of prior knowledge appears to be associated with the development of technologies that are less disruptive.

**Table 2** Regression models of disruptiveness and the use of prior knowledge and exaptation variables

|  | (1) | (2) | (3) | (4) | (5) | (6) |
|---|---|---|---|---|---|---|
| Minimum exaptation | 0.0739*** (0.0011) | 0.0764*** (0.0011) |  |  |  |  |
| Maximum exaptation |  |  | 0.0090*** (0.0007) | 0.0120*** (0.0007) |  |  |
| Average exaptation |  |  |  |  | 0.0395*** (0.0009) | 0.0435*** (0.0009) |
| Content similarity dispersion | -0.0728* | -0.0787*** | -0.0792*** | -0.0855*** | -0.0783*** | -0.0843*** |
| | ** | | | | | |
| | (0.0005) | (0.0005) | (0.0005) | (0.0005) | (0.0005) | (0.0005) |
| Field distance dispersion | -0.0372*** | -0.0360*** | -0.0510*** | -0.0509*** | -0.0492*** | -0.0485*** |
| | (0.0003) | (0.0003) | (0.0003) | (0.0003) | (0.0003) | (0.0003) |
| Ratio of self-citations to total work cited | -0.0409*** | -0.0475*** | -0.0423*** | -0.0486*** | -0.0417*** | -0.0482*** |
| | (0.0006) | (0.0006) | (0.0006) | (0.0006) | (0.0006) | (0.0006) |
| Mean age of work cited | 0.0019*** | 0.0004*** | 0.0020*** | 0.0005*** | 0.0019*** | 0.0004*** |
| | (0.0000) | (0.0000) | (0.0000) | (0.0000) | (0.0000) | (0.0000) |
| Dispersion in age of work cited | -0.0037*** | -0.0045*** | -0.0037*** | -0.0045*** | -0.0037*** | -0.0045*** |
| | (0.0000) | (0.0000) | (0.0000) | (0.0000) | (0.0000) | (0.0000) |
| Mean age of work cited × Dispersion in age of work cited | 0.0001*** | 0.0001*** | 0.0001*** | 0.0001*** | 0.0001*** | 0.0001*** |
| | (0.0000) | (0.0000) | (0.0000) | (0.0000) | (0.0000) | (0.0000) |
| Mean number of prior works produced by team members | 0.0000*** | 0.0000*** | 0.0000*** | 0.0000*** | 0.0000*** | 0.0000*** |
| | (0.0000) | (0.0000) | (0.0000) | (0.0000) | (0.0000) | (0.0000) |
| Year fixed effects | Yes | No | Yes | No | Yes | No |
| Field fixed effects | Yes | No | Yes | No | Yes | No |
| N | 3420673 | 3420673 | 3420673 | 3420673 | 3420673 | 3420673 |
| R2 | 0.12 | 0.10 | 0.11 | 0.10 | 0.12 | 0.10 |

*Notes:* This table evaluates the relationship between different measures of the use of prior technological knowledge and $CD_5$. Estimates are from OLS regressions. Each coefficient is tested against the null hypothesis of being equal to 0 using a two-sided t-test. We do not adjust for multiple hypothesis testing. Robust standard errors are shown in parentheses. Standard errors are in parentheses. + p<0.1; * p<0.05; ** p<0.01; *** p<0.001.

4.4 Robustness checks

To evaluate the robustness of our results, we introduced an exaptation dummy variable. This variable shifts the analysis from a quantitative to a qualitative perspective regarding exaptation-driven innovation. The dummy variable is binary, distinguishing values based on a predetermined threshold derived from the $E_{min}$ value. We set the upper 10%, 20%, and 30% of $E_{min}$ values as the respective thresholds. Instances with an $E_{min}$ exceeding these thresholds receive a dummy value of 1, which is indicative of exaptation-driven innovation, whereas those falling below are assigned a value of 0. This method is crucial, as not every patent stems from exaptation-driven innovation. Utilizing this dummy variable allows us to ascertain if the influence of exaptation on innovation correlates with increased disruptiveness. The proportion of exaptation-driven innovation exhibits a downward trend (as shown in Fig. 6b) across the various thresholds.

When establishing thresholds for the three groups at the upper 10%, 20%, and 30% threshold of $E_{min}$, the corresponding $E_{min}$ values are 0.281, 0.213, and 0.163, respectively. The square roots of these $E_{min}$ values are approximately 0.530, 0.461, and 0.404. This indicates that at a 10% threshold, the $E_{min}$ for the selected samples will surpass 0.281. Furthermore, given that content similarity and field distance both range from 0 to 1, their values in relation to $E_{min}$ will not drop below its square root, which is 0.530 for the 10% threshold. This rationale is consistent across the other two thresholds. The counts of the samples classified as exaptation-driven innovation at these thresholds are 342,068, 684,135, and 1,026,202, respectively.

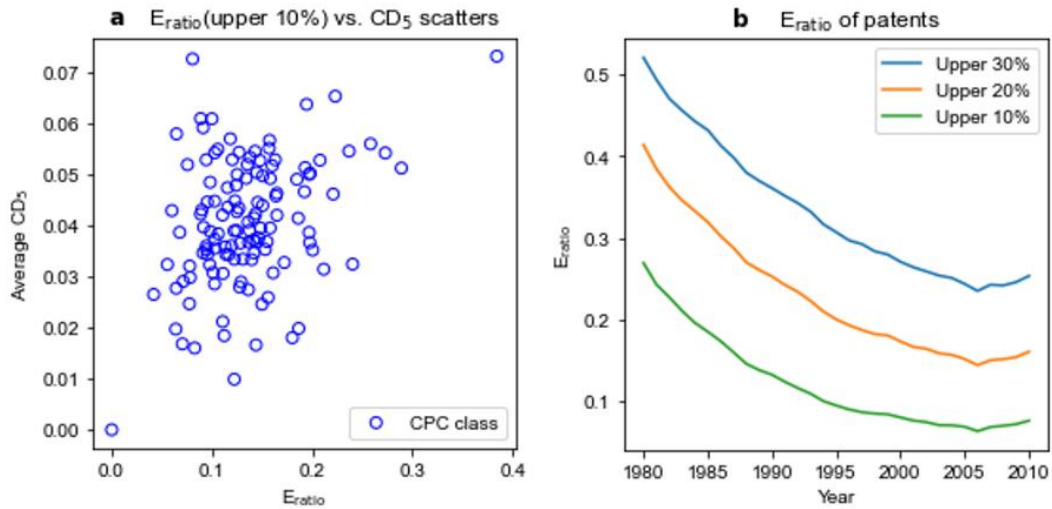

**Fig. 6** The scatters and trends of the exaptation ratio and trends of the exaptation ratio of patents

*Note:* **a**, the scatter plot displays the scatters of the average $CD_5$ and the ratio of exaptation-driven innovations, categorized by the upper 10% threshold at the CPC class level. The $E_{ratio}$ denotes the proportion of exaptation innovations relative to the total number of patents within a specific year. **b**, the chart illustrates the annual trends of the $E_{ratio}$ across various thresholds.

We employ multidimensional fixed effects regression models to examine the impact of exaptation on innovation disruptiveness (Table 3) and evaluate the outcomes with and without the absorption of year and field fixed effects. Models 1 and 2 analyze the regression results for the exaptation dummy with a 30% $E_{min}$ threshold. Models 3 and 4 examine the results for a 20% $E_{min}$ threshold and Models 5 and 6 for a 10% $E_{min}$ threshold. In line with prior regressions, all control variables yield similar outcomes. The positive and significant coefficient values of the exaptation dummy across the six models affirm that exaptation fosters innovation disruptiveness. Furthermore, as the $E_{min}$ threshold rises from 0.163 (upper 30% $E_{min}$) to 0.281 (10%), the exaptation dummy's value also increases. This indicates that higher baselines for identifying exaptation correlate with superior quality of exaptation and a stronger promotion of innovation disruptiveness. A positive correlation between disruptiveness and exaptation (upper 10%) is evident at the CPC class level, as depicted in Fig. 6a. Within each threshold group, the exaptation dummy's coefficient is higher when fixed effects are not absorbed compared to when they are, but R-square is lower. This observation suggests that the fixed effects may be accounting for some of the variability in disruptiveness that is otherwise captured by the exaptation dummy when fixed effects are not included. The higher coefficient without fixed effects implies that there may be unobserved factors, correlated with both exaptation and disruptiveness, that are not being controlled for in the fixed effects model. Therefore, while the exaptation dummy remains a significant predictor of innovation disruptiveness, the magnitude of its effect is somewhat attenuated when controlling for year and field-specific variations. This highlights the importance of considering these fixed effects to avoid overstating the impact of exaptation on disruptiveness.

**Table 3** |Regression models of disruptiveness and the use of prior knowledge and exaptation dummy

|  | Upper 30% | | Upper 20% | | Upper 10% | |
|---|---|---|---|---|---|---|
|  | (1) | (2) | (3) | (4) | (5) | (6) |
| Exaptation dummy | 0.0089*** (0.0002) | 0.0097*** (0.0002) | 0.0122*** (0.0002) | 0.0131*** (0.0002) | 0.0205*** (0.0004) | 0.0214*** (0.0004) |
| Content similarity dispersion | -0.0776*** (0.0005) | -0.0835*** (0.0005) | -0.0765*** (0.0005) | -0.0824*** (0.0005) | -0.0752*** (0.0005) | -0.0812*** (0.0005) |
| Field distance dispersion | -0.0433*** (0.0003) | -0.0420*** (0.0003) | -0.0431*** (0.0003) | -0.0419*** (0.0003) | -0.0438*** (0.0003) | -0.0428*** (0.0003) |
| Ratio of self-citations to total work cited | -0.0419*** (0.0006) | -0.0484*** (0.0006) | -0.0421*** (0.0006) | -0.0486*** (0.0006) | -0.0429*** (0.0006) | -0.0494*** (0.0006) |
| Mean age of work cited | 0.0019*** (0.0000) | 0.0004*** (0.0000) | 0.0019*** (0.0000) | 0.0004*** (0.0000) | 0.0019*** (0.0000) | 0.0004*** (0.0000) |
| Dispersion in age of work cited | -0.0036*** (0.0000) | -0.0045*** (0.0000) | -0.0036*** (0.0000) | -0.0044*** (0.0000) | -0.0036*** (0.0000) | -0.0044*** (0.0000) |
| Mean age of work cited × Dispersion in age of work cited | 0.0001*** (0.0000) | 0.0001*** (0.0000) | 0.0001*** (0.0000) | 0.0001*** (0.0000) | 0.0001*** (0.0000) | 0.0001*** (0.0000) |
| Mean number of prior works produced by team members | 0.0000*** (0.0000) | 0.0000*** (0.0000) | 0.0000*** (0.0000) | 0.0000*** (0.0000) | 0.0000*** (0.0000) | 0.0000*** (0.0000) |
| Year fixed effects | Yes | No | Yes | No | Yes | No |
| Field fixed effects | Yes | No | Yes | No | Yes | No |
| N | 3420673 | 3420673 | 3420673 | 3420673 | 3420673 | 3420673 |
| R2 | 0.12 | 0.10 | 0.12 | 0.10 | 0.12 | 0.10 |

*Notes:* This table evaluates the relationship between different measures of the use of prior technological knowledge and $CD_5$. Estimates are from OLS regressions. Each coefficient is tested against the null hypothesis of being equal to 0 using a two-sided t-test. We do not adjust for multiple hypothesis testing. Robust standard errors are shown in parentheses. Standard errors in parentheses. + $p<0.1$; * $p<0.05$; ** $p<0.01$; *** $p<0.001$.

## 5. Discussion

This research empirically validates the connection between exaptation-driven innovation and disruptive innovation. The CD index is a significant gauge for evaluating whether an innovation possesses disruptiveness, as posited by Funk and Owen-Smith (2017); however, it is a trailing indicator that cannot preemptively determine if an innovation will cause disruption based on its inherent attributes. When considering technology exaptation, it is still unclear whether the technologies of a previous generation will be referenced in future works across different fields. Therefore, the concept of technology exaptation is often intertwined with serendipity (Garud et al., 2018; Leporini et al., 2020). Yet, through the use of content similarity and field distance, we can establish exaptation values to assess how much exaptive knowledge a focal patent has derived from its predecessors, thus gauging whether an innovation is driven by exaptation. The exaptation values are proactive indicators, and our regression results indicate they have a facilitative effect on disruptiveness. This provides guidance on how to more effectively develop disruptive innovation.

While our findings align with other studies that suggest disruption decreases over time (Park et al., 2023), they also reveal that the technological community's continuous attempts with exaptation is on the rise, as indicated by the annual growth trend of Emax (Fig. 3b). The phenomenon of inventors consistently learning from other domains and applying these technologies to their own fields is on the increase, signifying technology diffusion, even though the proportion of exaptation-driven innovation is in decline (Fig. 6b).

From $E_{max}$, we cannot distinctly discern the promotion of disruption by exaptation, since a patent often cites numerous others, and the exaptive knowledge within these patents may only represent a small fraction of the knowledge. Such patents cannot be classified as exaptation-driven innovations but are rather patents that have incorporated some exaptive knowledge. Therefore, focusing solely on $E_{max}$ could lead to a substantial deviation. In contrast, $E_{min}$, as the baseline for exaptation, is indeed a robust indicator for assessing exaptation, with regression results showing a strong positive correlation with disruption. Setting a threshold for $E_{min}$ can also assist in determining whether a patent is an exaptation-driven innovation. Furthermore, exaptation not only enhances disruptiveness within a five-year span but also contributes to long-term disruption growth, as evidenced by the positive correlation between the variable ΔCD and $E_{min}$.

Although inventors' attempts at exaptation are increasing, the ratio of genuine exaptation-driven innovations is declining. To make innovation more disruptive, it is insufficient to merely borrow knowledge from other fields; it must evolve into exaptation-driven innovation. The prior knowledge utilized must be concentrated on specific technologies and domains, not overly diverse, to avoid forming a less disruptive, confusing hodgepodge of technologies that may lack the cohesion necessary for disruptiveness. From the perspective of exaptation's impact on disruption, for disruptive innovation, inventors need to metaphorically stand on the shoulders of giants and venture into different fields, thereby fostering innovation that surpasses initial intentions.

Given that this study's data is limited to USPTO patent data from 1980 to 2010, it is unclear if innovations from other years and fields follow similar patterns. Further research is required to ascertain whether exaptation also promotes disruptiveness in the progeny of focal patents.

# Supplementary materials

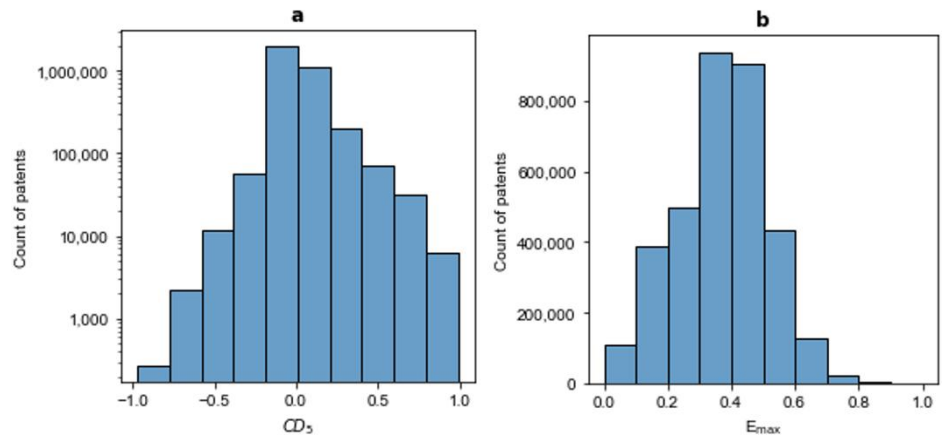

**Extended Data Fig. 1** Distribution of CD5 and E$_{max}$

This figure provides an overview of the distribution of CD5 and Emax for patents (n = 3,420,673).

**Extended Data Table 1** Illustrative patents

| Patent | Grant year | $CD_5$ | $\triangle CD$ | $E_{max}$ | $E_{min}$ | $CS_{disp}$ | $FD_{disp}$ | $Impact_5$ | Ancestor count | Title | Assignee |
|---|---|---|---|---|---|---|---|---|---|---|---|
| 4,356,429 | 1982 | 0 | 0.2 | 0.314 | 0.19 | 0.233 | 0.129 | 2 | 2 | Organic electroluminescent cell | Eastman Kodak Company |
| 4,399,216 | 1983 | 0.8 | 0.067 | 0.327 | 0.327 | 0 | 0 | 15 | 1 | Processes for inserting DNA into eucaryotic cells and for producing proteinaceous materials | Columbia University |
| 4,445,050 | 1984 | 0.167 | 0.141 | 0.502 | 0.336 | 0.093 | 0.175 | 1 | 2 | Device for the conversion of light power to electric power | None |
| 4,573,530 | 1986 | 0.2 | -0.033 | 0.484 | 0 | 0.104 | 0.75 | 1 | 2 | In-situ gasification of tar sands utilizing a combustible gas | Mobil Oil Corporation |
| 4,637,464 | 1987 | 0 | -0.1 | 0.416 | 0.119 | 0.351 | 0.483 | 0 | 7 | In situ retortion of oil shale with pulsed water purge | Amoco and Chevron |
| 4,658,215 | 1987 | 0.222 | -0.04 | 0.054 | 0 | 0.568 | 0.125 | 2 | 3 | Method for induced polarization logging | Shell Oil |
| 4,724,318 | 1988 | 0.027 | 0.038 | 0.371 | 0.334 | 0.046 | 0.125 | 24 | 2 | Atomic force microscope and method for imaging surfaces with atomic resolution | IBM |
| 4,928,765 | 1990 | 0 | 0 | 0.374 | 0.049 | 0.079 | 0.542 | 0 | 6 | Method and apparatus for shale gas recovery | Ramex Syn-Fuels |
| 5,010,405 | 1991 | 0.45 | 0.273 | 0.175 | 0.145 | 0.122 | 0 | 9 | 2 | Receiver-compatible enhanced definition television system | MIT |
| 5,015,744 | 1991 | -0.491 | 0.102 | 0.488 | 0.218 | 0.206 | 0.265 | 32 | 4 | Method for the preparation of taxol using an oxazinone | Florida State University |
| 5,016,107 | 1991 | 0.029 | 0.069 | 0.432 | 0.157 | 0.263 | 0.5 | 19 | 17 | Electronic still camera utilizing image compression and digital storage | Eastman Kodak Company |
| 6,063,738 | 2000 | 0 | -0.168 | 0.438 | 0.231 | 0.406 | 0.417 | 41 | 11 | Foamed cement slurries, additives and methods | Halliburton Energy Services, Inc. |
| 6,285,999 | 2001 | 0.159 | 0.23 | 0.326 | 0.099 | 0.358 | 0.542 | 31 | 7 | Method for node ranking in a linked database | Stanford University |

| | | | | | | | | | | |
|---|---|---|---|---|---|---|---|---|---|---|
| 6,376,284 | 2002 | -0.266 | 0.071 | 0.585 | 0.155 | 0.383 | 0.681 | 93 | 19 | Method of fabricating a memory device | Micron Technology, Inc. |
| 6,958,436 | 2005 | -0.843 | -0.075 | 0.428 | 0 | 0.534 | 0.458 | 149 | 5 | Soybean variety SE90346 | Monsanto |